\documentclass[prb,twocolumn,showpacs,amssymb]{revtex4}

\usepackage{graphicx}
\usepackage{dcolumn}
\usepackage{bm}

\begin{document}


\title{Theory of Excitonic States in Medium-Sized Quantum Dots}
\author{Alexander Odriazola$^a$, Alain Delgado$^b$, Augusto Gonzalez$^a$}
\affiliation{$^a$ Instituto de Cibernetica, Matematica
 y Fisica, Calle E 309, Vedado, Ciudad Habana, Cuba\\
 $^b$ Centro de Aplicaciones Tecnologicas y Desarrollo Nuclear,
 Calle 30 No 502, Miramar,Ciudad Habana, C.P. 11300, Cuba}

\begin{abstract}
In a quantum dot with dozens of electrons, an approximation beyond Tamm-Dankoff is 
used to construct the quantum states with an additional electron-hole pair, 
i.e. the ``excitonic'' states. The lowest states mimic the non-interacting 
spectrum, but with excitation gaps renormalized by Coulomb interactions. At 
higher excitation energies, the computed density of energy levels shows an 
exponential increase with energy. In the interband absorption, we found a
background level in the quasicontinuum of states rising linearly with the 
excitation energy. Above this 
background, there are distinct peaks related to single resonances or to groups
of many states with small interband dipole moments.
\end{abstract}


\maketitle

\section{Introduction}

Semiconductor quantum dots are like anisotropic Thomson atoms. Most of their 
electronic and optical properties are, by now, well studied \cite{QDots_Review}. 
The typical energy scales determining these properties are the following: the 
in-plane confinement strength, $\hbar\omega$, which also defines the unit of 
length, the characteristic Coulomb energy, $\beta$, and the characteristic 
energy along the symmetry axis, $E_{z}$. The ratio $\beta/\hbar\omega$ is an 
indicator of the role played by Coulomb interactions in the dot. In the strong 
confinement regime, $\beta/\hbar\omega\ll 1$, which we encounter in 
self-assembled structures, quantum dots are like real atoms, with correlation 
energies reaching only a few percents of the total energy. On the other hand, 
for $\beta/\hbar\omega\gg 1$, correlations are strong, and there could be even 
premonitions of  Wigner crystalization for large enough dots.

The excited states of a quantum dot, as in most quantum systems, subscribe to a 
general picture of a few low-lying ``oscillator'' states followed by a 
quasi-continuum of excitations. Typically, the low lying states determine most 
of the optical properties. But there are situations in which higher excitations 
play a key role. Raman scattering or PLE experiments with excitation energies 
$40-60$ meV above band gap, for instance, sense high interband excitations. On 
the other hand, the density of energy levels in the quasi-continuum obeys a 
simple exponential rule with excitation energy, known in Nuclear Physics as 
the Constant Temperature Approximation (CTA) \cite{CTA_in_Nucl_Phys}. 
In paper [\onlinecite{PRB_con_Capote}], we corroborated this picture for the 
intraband excitations of few-electron quantum dots in a magnetic field.

In the present paper, we focus on the interband excitations. A medium-size dot, 
charged with $42$ electrons, under intermediate confinement regime 
($\beta/\hbar\omega\approx0.6$) is studied, a situation typical of etched dots. 
As it will be shown below, the obtained low-lying spectrum is similar to the 
spectrum of the non-interacting model, but with values renormalized by Coulomb 
interactions. The quasi-continuum of states follows the CTA, as expected.

\begin{figure}[t]
\begin{center}
\includegraphics[width=.95\linewidth,angle=0]{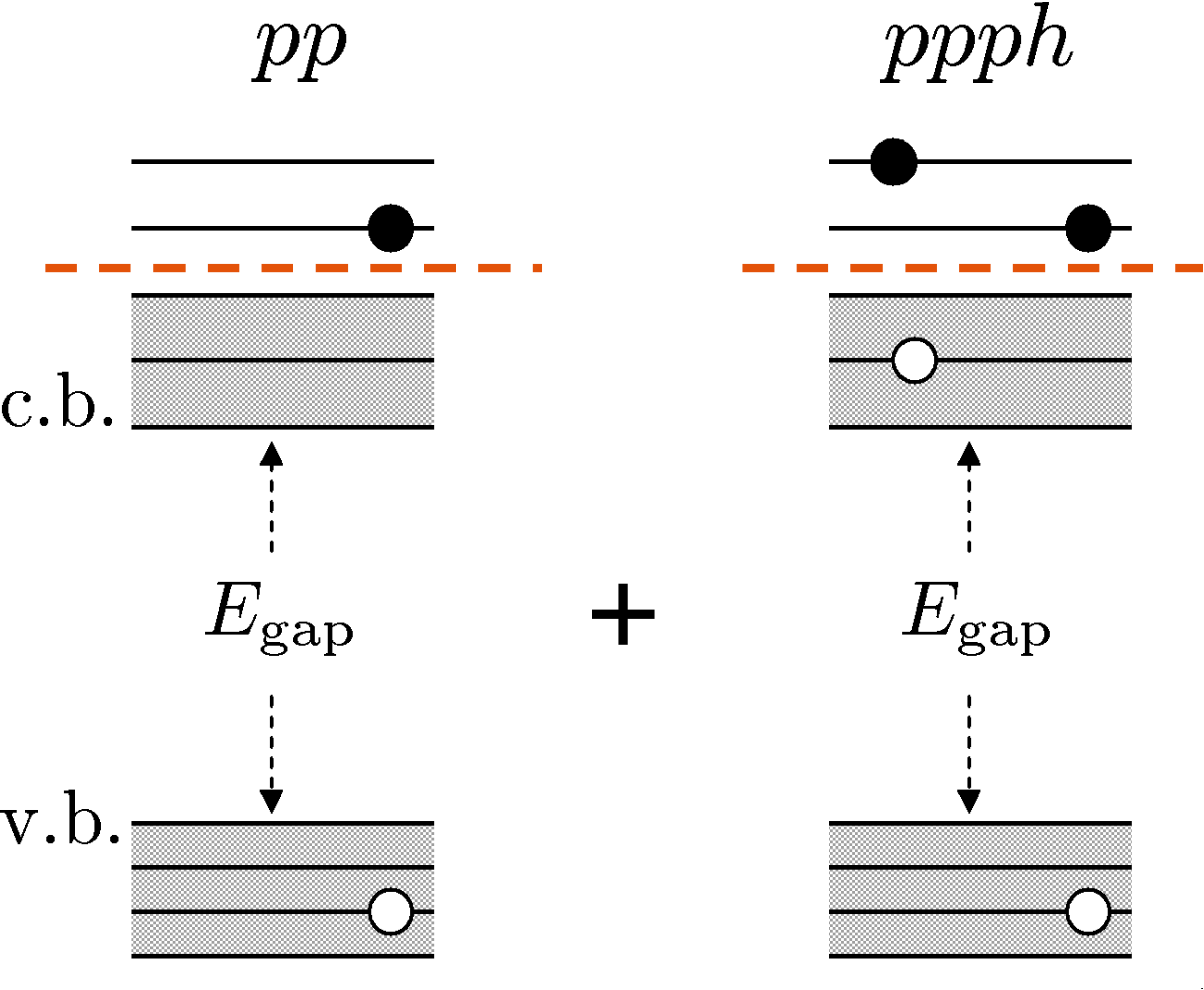}
\caption{\label{fig1} (Color online) The pp-TDA and ppph contributions to the 
wave function of the excitonic states.}
\end{center}
\end{figure}

For a dot with $42$ electrons, no exact diagonalization calculations are 
possible and we should resort to approximations. We will use a kind of 
Quantum Chemistry Configuration Interaction Method (CI) \cite{C_I} in which 
the starting point is the Hartree-Fock (HF) solution for the $42$-electron 
problem. Excitations over the HF solution conform the basis functions in which 
our Hamiltonian is to be diagonalized. The simplest set of excitations, one 
additional electron in the conduction band (CB) and one hole in the valence 
band (VB), is usually called pp-Tamm-Dankoff approximation (pp-TDA) in Nuclear 
Physics \cite{Ring}. The valence-band hole is treated as a quasiparticle with
its own dispersion relation. Excitonic states are, thus, states with two
additional ``particles'' above the HF solution. The next set in increasing 
complexity includes excitations with an additional electron above the Fermi 
level and a hole below the Fermi level in the CB. We will call it the ppph 
contribution. A schematic representation of the pp-TDA and ppph contributions
to the wave function of the excitonic states is given in Fig. \ref{fig1}. 
We will truncate the basis of functions at this stage.

The reason for including next-to-leading order excitations in the basis 
functions, i.e. going beyond pp-TDA, is the following. It is well known that
the pp-TDA underestimates the density of energy levels already for relatively
low excitation energies. A correct level density could be crucial in the
description of Raman scattering events with laser excitation energy well 
above band gap, which is our next goal. A particularly interesting question is
why the so called outgoing resonances, presumably a result of higher order 
Raman processes \cite{Danan_&_Pinczuk}, give stronger contribution to the 
Raman spectrum than the usual second-order Raman amplitude.

\section{Formalism}
\label{sec2}

As mentioned above, the starting point in our formalism is the HF solution for 
the $42$-electron problem. The HF equations and the way we solve them are 
described elsewhere \cite{Alain_PRB, Alain_thesis}. We use the following set of 
parameters: $m_{e}=0.067~m_{0}$, (a GaAs quantum dot), $\hbar\omega_{e}=12$ meV 
(the confinement energy for electrons), and $\beta=7.07$ meV. The latter value 
comes from the expression: $\beta=0.6~e^2/(4\pi\epsilon_0\kappa l)$, where the 
relative dielectric constant is $\kappa=12.8$, $l$ is the oscillator length, and
the factor 0.6 takes account of the quasibidimensionality of the structure. 
Only one electron 
sub-band in the z-direction is considered, and the value of $E_{z}^{(e)}$ is 
used to control the effective band gap of the nanostructure.

The HF states for the hole in the VB are found from the Kohn-Luttinger 
Hamiltonian, in which a term accounting for the background potential created by 
the $42$ electrons is included \cite{Alain_PRB, Alain_thesis}.

For the interband excitations of the dot, i.e. many-particle wave functions 
with $43$ electrons and one hole, we write the ansatz,

\begin{eqnarray}
\vert int \rangle&=&\left (\sum_{\sigma,\tau}V_{\sigma\tau} 
 e^\dagger_\sigma h^\dagger_\tau + 
\sum_{\sigma,\rho,\lambda,\tau}Z_{\sigma\rho\lambda\tau} 
 e^\dagger_\sigma e^\dagger_\rho e_\lambda h^\dagger_\tau\right )
 \vert HF\rangle\nonumber\\ 
&=& Q^\dagger\vert HF\rangle.
\label{Eq.Int.}
\end{eqnarray}

\noindent 
$\sigma$ and $\rho$ are electronic states above the Fermi level, 
and $\lambda$ is a state below the Fermi level of electrons. We restrict the 
second sum in Eq. (\ref{Eq.Int.}) to $\sigma>\rho$ in order to avoid 
overcounting. This second sum is the term beyond pp-TDA, not included in 
previous computations of interband excitations \cite{Alain_PRB, Alain_thesis}.

Due to the peculiarities of Coulomb interactions, the Kohn-Luttinger 
Hamiltonian, and the central confinement potential, the interband excitations 
are characterized by two quantum numbers, $F$ and $S_{z}$.  It means that 
the single-particle states entering the first and second sums of Eq. 
(\ref{Eq.Int.}) should satisfy, respectively, the first and second rows of the 
following equalities:

\begin{eqnarray}
F&=&l_{\sigma}+l_{\tau}-m_j\nonumber\\ 
&=&l_{\sigma}+l_{\rho}-l_{\lambda}+l_{\tau}-m_j,
\label{F}
\end{eqnarray}

\begin{eqnarray}
S_{z}&=&S_{\sigma}\nonumber\\ 
&=&S_{\sigma}+S_{\rho}-S_{\lambda}.
\label{Sz}
\end{eqnarray}

\noindent 
Here $l$ are (orbital) angular momentum, $m_j$ - (band) hole angular 
momentum, and $S$ - (electron) spin quantum numbers. Of course, these 
magnitudes correspond to projections along the $z$-axis.

The coefficients $V$ and $Z$ (the wave functions) and the excitation energies 
with respect to the HF state are computed from the following eigenvalue problem, 
in close analogy with the TDA \cite{Ring}

\begin{equation}
[H, Q^\dagger]\vert HF\rangle=\Delta E_{int}~ Q^\dagger\vert HF\rangle,
\label{Eq.Mov.}
\end{equation}

\noindent 
where $H$ is the many-particle electron-hole Hamiltonian

\begin{equation}
H=T_e+T_h+V_{ee}+V_{eh}.
\end{equation}

\noindent
The $T$ are one-body operators, and $V$ refer to Coulomb interactions. Let us
stress that electron-hole exchange, usually of the order of $\mu$eV in these
structures, is neglected. The explicit matrix form 
of Eq. (\ref{Eq.Mov.}) is presented in Appendix A for completeness. We use 
energy cutoffs of 90 meV to reduce the dimensions of the resulting matrices 
to around 12000. Eigenvalues and eigenvectors (the wave functions) are obtained 
by numerical diagonalization.

\section{The lowest excitonic states}
\label{sec3}

The solution of Eq. (\ref{Eq.Mov.}) provides us the wave functions and energies 
of the interband excitations of the dot (the excitonic states). In this section 
we focus on the lowest states.

\begin{figure}[t]
\begin{center}
\includegraphics[width=.98\linewidth,angle=0]{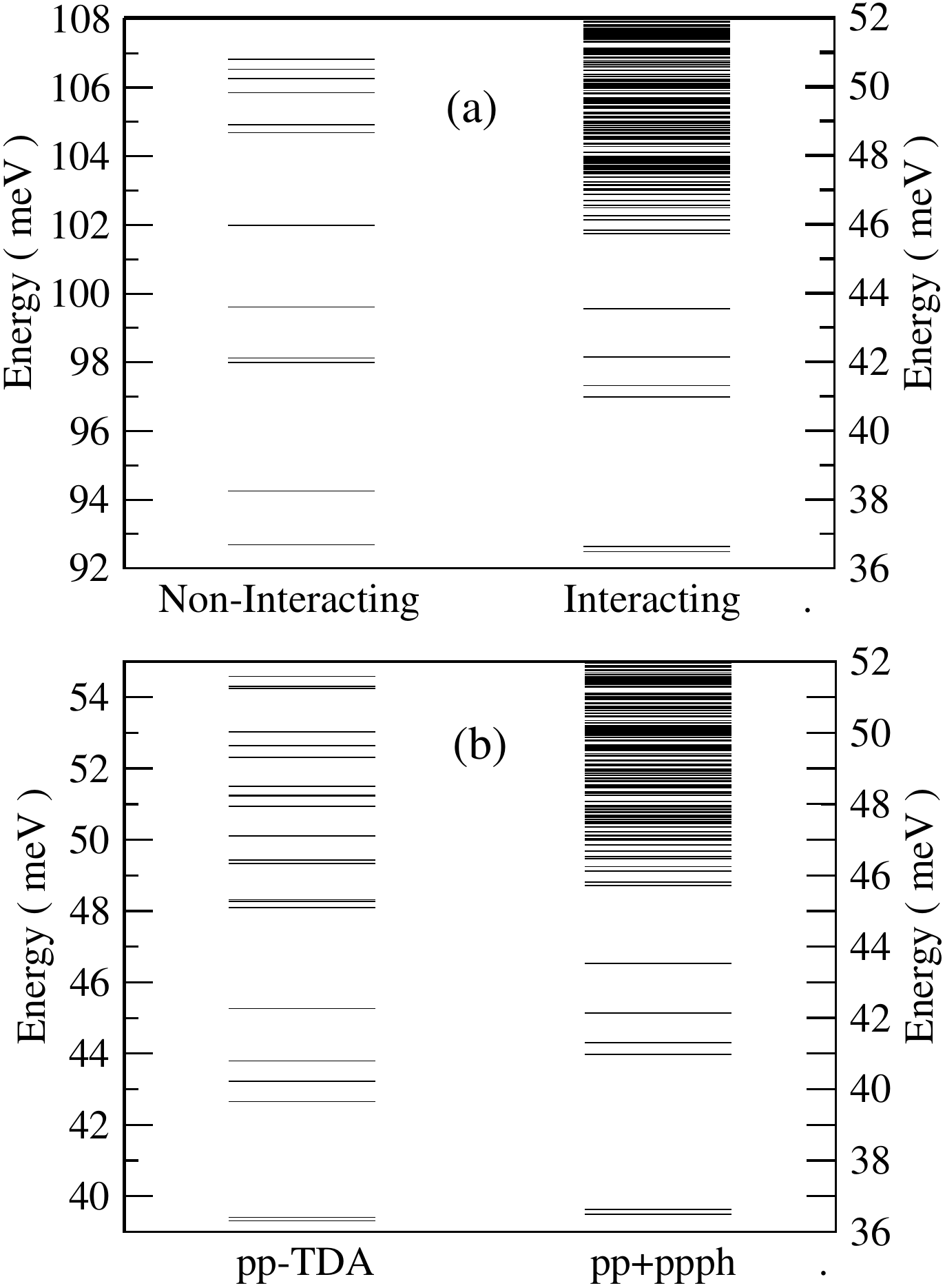}
\caption{\label{fig2} (a) The low-lying interband excitations in the model 
quantum dot. States with $F=-3/2$, $S_z=1/2$  are drawn. Results from the 
non-interacting picture are also shown for comparison. (b) Increase in the
density of energy levels due to the ppph component.}
\end{center}
\end{figure}

A sample of results is shown in Fig. \ref{fig2}(a). Energies are measured with 
respect to $E_{HF}+E_{z}^{(e)}$. States with quantum numbers $F=-3/2$, $S_z=1/2$ 
are represented in the figure. With increasing energy we observe first a pair 
of states followed by a group of four states, and then a quasi-continuum of 
excitations. This sequence can be understood in terms of the non-interacting 
picture of electrons and holes.

Our model dot with $42$ electrons has $6$ filled $2D$ oscillator shells. We 
should add an electron-hole pair to construct an interband excitation. The 
electron should then be added to the 7th shell. The possible angular momentum 
values are $l_{e}=-6,-4,\dots,6$. The hole orbital state, on his hand, can be 
the first oscillator state with $l_{h}=0$. As we are talking about $F=-3/2$ 
states, only two combinations remain: $l_{e}=0$, $mj=3/2$ and $l_{e}=-2$, 
$mj=-1/2$. That is, one heavy and one ligth hole exciton that, because of the 
assumed $25$ nm width of the dot in the $z$-direction, are very close in energy 
(1.5 meV). The next set of 4 states are related to hole excitations, i.e. the 
hole occupies one oscillator state of the 2nd shell. We choose the confinement 
frequency for holes in such a way that the oscillator length is unique, meaning 
that electrons and holes are confined in the same spatial region. For heavy 
holes, for example, it means that $\omega_{hh}=m_{e}\omega_{e}/m_{hh}$. As a 
result, excitations energies for holes are smaller, of around 5 meV. Next, 
we reach a group, more numerous, of states built up from one-$\hbar\omega$ 
electron excitations (12 meV) or two-$\hbar\omega$ hole excitations.

Following this non-interacting picture, we can assign the first two levels to 
heavy-hole and ligth-hole excitons, the next four to hole excitations to the 
2nd oscillator shell, and we can relate the quasi-continuum of states to 
electron excitations to the 8th shell and hole excitations to the third shell. 
The effects of Coulomb interactions are apparent in Fig. \ref{fig2}(a): energy 
values are pushed down, energy gaps are reduced and a quasi-continuum of 
levels appear at excitation energies even below $\hbar\omega_e=12$ meV. 

In Fig. \ref{fig2}(b) we illustrate the main effects of the ppph component on
the energy levels. As compared with the pp-TDA, we get small variations of
the ground state energy (of around 3 meV) and the gap to the first quartet of 
states. But, for excitation energies above 10 meV a significant increase in
the level density is observed. In the next section a more detailed study of
the density of energy levels is carried on. 

\section{The Constant Temperature Approximation (CTA)}
\label{sec4}

It can be verified that the CTA is a good approximation for the density of 
energy levels at relatively small excitation energies, starting from the 
quasi-continuum of states.

\begin{figure}[t]
\begin{center}
\includegraphics[width=.98\linewidth,angle=0]{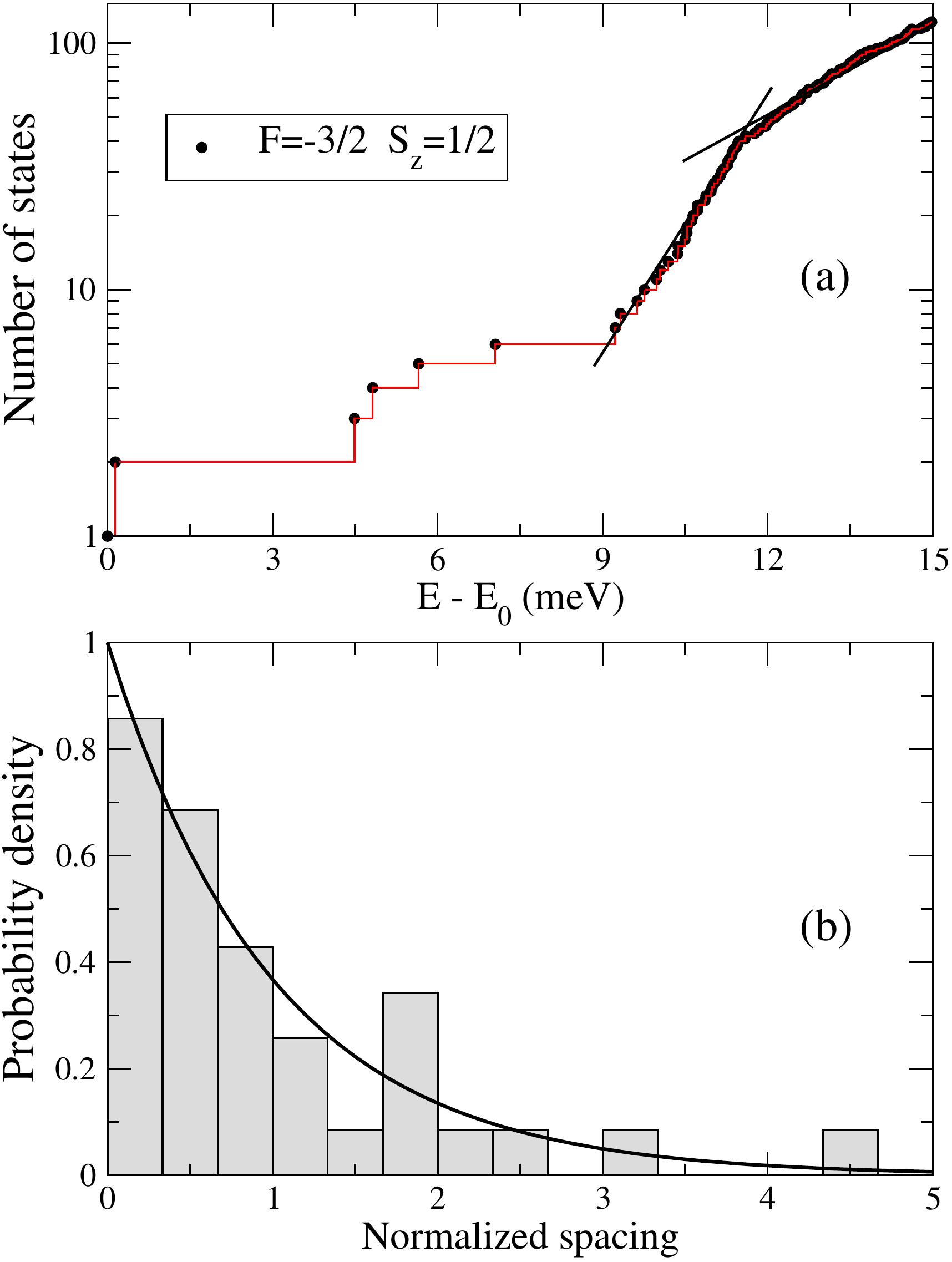}
\caption{\label{fig3} (Color online) (a) Increase in the number of levels as 
predicted by the CTA for states with 9 meV $< \Delta E <$ 15 meV. (b) 
Probability distribution of level spacing for excitonic states in the interval 
9 meV $< \Delta E <$ 12 meV.}
\end{center}
\end{figure}

We show in Fig. \ref{fig3}(a) the excitonic sector with quantum numbers 
$F=-3/2$, $S_z=1/2$. Excitation energies, measured with respect to the lowest 
state in this sector, run in the interval 0 - 15 meV. A perfect exponential 
behaviour is apparent in the regions 9 meV $< \Delta E <$ 12 meV and 12 meV 
$< \Delta E <$ 15 meV. In these intervals, the number of states can be fitted 
to:

\begin{equation}
N=N_{0}\exp\left(\frac{\Delta E}{\Theta}\right),
\label{CTA}
\end{equation}

\noindent where: $N_{0}=4.472\times10^{-3}$, $\Theta=1.268$ meV for 9 meV 
$< \Delta E <$ 12 meV, and $N_{0}=1.145$, $\Theta=3.183$ meV for 12 meV 
$< \Delta E <$ 15 meV. 

We guess that a simple qualitative relation should exist between the 
temperature parameter, $\Theta$ (the slope), and the system parameters. Indeed,
in a scaled formulation \cite{GPP} $\Theta$ should be proportional to 
$\hbar\omega_e$, simply because of dimensional arguments. The number of 
electrons, $N_e$, and the scaled Coulomb strength, 
$\hat\beta=\beta/(\hbar\omega_e)$ 
should appear in the combination $N_e^{1/4}\hat\beta$.\cite{GPP} We expect the
following dependence: $\Theta\sim \hbar\omega_e/(N_e^{1/4}\hat\beta)^{2/3}$.
A verification of the above dependence requires extensive calculations and is
delayed for a future work.

The slope discontinuity at $\Delta E\approx 12$ meV is a reminiscence of shell 
structure. Indeed, if there were a gap between perfect shells, then the curve 
would be flat immediatly after $\Delta E\approx 12$ meV. Coulomb interaction 
is responsible for filling the ``gap'' region, 12 meV $< \Delta E <$ 15 meV, 
with levels.

The interpretation of levels in the interval 9 meV $< \Delta E <$ 12 meV as a 
``distorted shell'' is also reinforced by the distribution of spacing between 
neighbourgh levels, as shown in Fig. \ref{fig3}(b). The 
probability density vs spacing, normalized to the mean level spacing, which in 
the present situation is 0.067 meV, is shown to follow a Poissonian dependence, 
indicating a regular region of phase space \cite{Level_spacing_in_Q_Systems}.
This statement proves that the CTA is not necessarily related to chaos in the
excitation spectrum.

In conclusion, the quasi-continuum of excitonic states for 9 meV 
$< \Delta E <$ 12 meV $\approx\hbar\omega_{e}$ exhibits regular or 
quasi-integrable behaviour and is well described by the CTA. Above this
interval, an abrupt change in the temperature parameter is reminiscent of
shell strucure in the model.

\section{Interband Absorption}

Interband absorption at normal incidence is characterized by the creation of 
electron-hole pairs mainly in monopole states, $l_{e}+l_{h}=0$. This means low 
values of $\vert F\vert$.

The transition matrix element for absorption is given by

\begin{equation}
\langle int\vert H^{(-)}\vert i\rangle = \sum_{\sigma,\tau}
V_{\sigma\tau}^{*}b_{\sigma\tau},
\label{absorpt}
\end{equation}

\noindent where the initial state is the HF state for the $42$-electron system, 
and $H^{(-)}$ is the light-matter coupling Hamiltonian corresponding to the 
absorption of a photon and creation of a pair. An explicit expresion for 
$H^{(-)}$ is given in Refs. [\onlinecite{ORA,Alain_thesis}]. The matrix elements 
$b_{\sigma\tau}$ are explicitly given in Ref. [\onlinecite{Alain_PRB}], 
where they are termed band-orbital factors (see Eq. (30) of that paper). Notice 
that only the first part of the wave function (\ref{Eq.Int.}) enters the 
matrix element (\ref{absorpt}). This is the result of treating the ground state
in the Hartree-Fock approximation. 1p1h corrections would not modify Eq.
(\ref{absorpt}) because the Hartree-Fock state is orthogonal to 1p1h corrections
\cite{Ring}.

\begin{figure}[t]
\begin{center}
\includegraphics[width=.98\linewidth,angle=0]{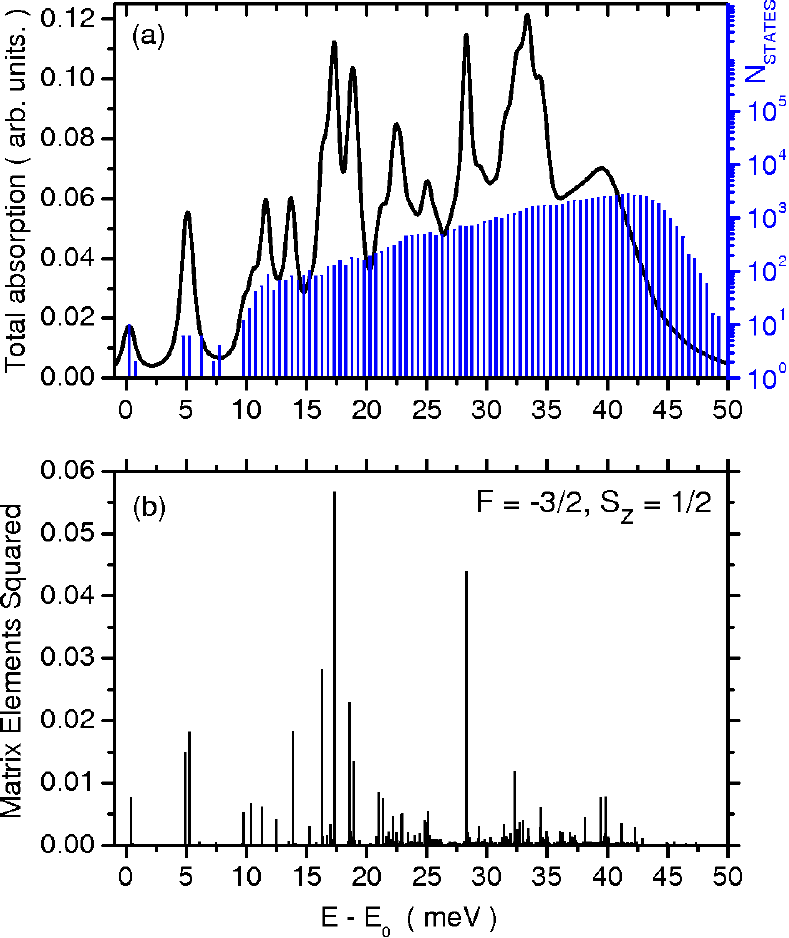}
\caption{\label{fig4}(Color online) (a) Interband absorption in our model dot
and the density of energy levels. 
(b) Contribution to the absorption by the set of states with quantum 
numbers $F=-3/2$, $S_z=1/2$.}
\end{center}
\end{figure}

We show in Fig. \ref{fig4}(a) the computed zero-temperature interband absorption 
in our model quantum dot. We consider normal incidence and linear polarization. 
The absorption intensity has contributions from four sets of states, that is
$(F,S_z)$=(-3/2,1/2), (3/2,-1/2), (-1/2,-1/2) and (1/2,1/2). A Lorentzian 
broadening of the spectral lines is assumed. 
The energy widths, $\Gamma_{int}$, are choosen in the following way: 
$\Gamma_{int}=0.5$ meV for $\Delta E < 35$ meV, and $\Gamma_{int}=2.0$ for 
$\Delta E > 35$ meV, where $\Delta E$ is the energy difference with respect to 
the first excitonic state, and the value 35 meV is the LO phonon energy in GaAs. 
Our ansatz for $\Gamma_{int}$ expresses the fact that the opening of the channel 
for the emission of a LO phonon broadens the excitonic levels.

The absorption spectrum has the usual form. First, a few isolated peaks, and 
then a quasicontinuum. 
We have superimposed in Fig. \ref{fig4}(a) a histogram with the density of
energy levels, in logarithmic scale. To construct the histogram, we used an 
energy width equal to $\Gamma_{int}$, that is 0.5 meV. The decrease of the
density of states for $\Delta E\ge 45$ meV is an artifact related to 
the truncation of Hilbert space. It becomes evident that in the
quasicontinuum there is a background absorption intensity roughly proportional 
to the logarithm of the level density, i.e. to the excitation energy $\Delta E$
(because of the CTA). We notice that this is the dependence that follows for
absorption in the independent-particle picture for electrons in a harmonic
potential. Above this background level, we observe distinct peaks. 
 
The contribution to the interband absorption by the set of states with quantum 
numbers $F=-3/2$, $S_z=1/2$ is sketched in Fig. \ref{fig4}(b). The comparison
with Fig.\ref{fig4}(a) reveals that there are peaks associated to one (or a
few) excitonic states with strong interband dipole moments, but there are also
peaks due to the contribution of many states with relatively small dipole 
moments. The latter are better understood as broad resonances, but they 
should be distinguished from the background intensity created by the 
``single-pair'' excitations.

\section{Concluding remarks}

In the present paper, we have computed the excitonic states in a relatively 
large quantum dot by using a Configuration Interation scheme. This is an
ab-initio procedure, fully variational. The only approximations in it are the
effective mass description of electrons, the Kohn-Luttinger description of
holes in the valence band and the harmonic model for the dot. To the best of
our knowledge, this is the first calculation of such kind in the literature.

For the sake of simplicity in the presentation, we used a closed-shell situation, 
42 electrons, in which the ground state is characterized by quantum numbers 
$L_{total}=0$, and $S=0$. However, our scheme works properly also
for open-shell dots and, with appropiate 2p2h corrections, provides quantum 
numbers for the ground state in accordance with Hund's rules\cite{2p2h}. Comparison
with exact calculations for few-electron dots shows excelent agreement \cite{2p2h}.

The main results of the paper are the following: (a) A picture for the low-lying
states in which the role of Coulomb interactions is made evident, (b) A 
universal parametrization of the density of energy levels for excitation 
energies greater than 1 $\hbar\omega$, and (c) A picture for the interband
absorption made up from peaks of different nature and a background level
in the quasicontinuum rising linearly with energy.

The work may be continued along different lines. A natural extension is the
description of biexcitonic states. The verification of 
universality of the CTA in the level density of quantum systems is also worth 
trying. Finally, the 
obtained excitonic states can be used in the computation of Raman cross 
sections for laser excitation energies well above band gap. Reasearch along 
these directions is currently in progress.

\begin{acknowledgments}
Part of this work was performed during a stay of A.D. at the Abdus Salam ICTP, 
Trieste, Italy. The authors 
acknowledge also support by the Caribbean Network for Quantum Mechanics,
Particles and Fields (ICTP-TWAS) and by the Programa Nacional de Ciencias
Basicas (Cuba).
\end{acknowledgments}
\vspace{.5cm}

\appendix

\section{Eq. (\ref{Eq.Mov.}) in matrix form}

Eq. (\ref{Eq.Int.}) for the $Q^\dagger$ operator makes evident the basis
functions used in the construction of the excitonic states. They are of two
kinds:

\begin{equation}
 e^\dagger_\sigma h^\dagger_\tau \vert HF\rangle,
\end{equation}

\noindent and

\begin{equation}
 e^\dagger_\sigma e^\dagger_\rho e_\lambda h^\dagger_\tau \vert HF\rangle.
\end{equation}

If Eq. (\ref{Eq.Mov.}) is projected onto these basis functions, we get the 
following system of equations:

\begin{widetext}
\begin{eqnarray}
\sum_{\sigma\tau}\langle HF\vert [h_{\tau^\prime} e_{\sigma^\prime},
 [H,e^\dagger_{\sigma} h^\dagger_{\tau}]]\vert HF\rangle V_{\sigma\tau}
 &+&\sum_{\sigma\rho\lambda\tau}\langle HF\vert [h_{\tau^\prime}
 e_{\sigma^\prime},[H,e^\dagger_{\sigma} e^\dagger_{\rho}
 e_{\lambda} h^\dagger_{\tau}]]\vert HF\rangle 
 Z_{\sigma\rho\lambda\tau}\nonumber\\
&=& \Delta E_{int} \sum_{\sigma\tau}\langle HF\vert [h_{\tau^\prime}
 e_{\sigma^\prime},e^\dagger_{\sigma} h^\dagger_{\tau}]\vert HF
 \rangle V_{\sigma\tau},
\end{eqnarray}

\begin{eqnarray}
\sum_{\sigma\tau}\langle HF\vert [h_{\tau^\prime} e^\dagger_{\lambda^\prime}
 e_{\rho^\prime} e_{\sigma^\prime},[H,e^\dagger_{\sigma} h^\dagger_{\tau}]]
 \vert HF\rangle V_{\sigma\tau}\ &+&
\sum_{\sigma\rho\lambda\tau}\langle HF\vert [h_{\tau^\prime} 
 e^\dagger_{\lambda^\prime} e_{\rho^\prime} e_{\sigma^\prime},[H,
  e^\dagger_{\sigma} e^\dagger_{\rho} e_{\lambda} h^\dagger_{\tau}]]\vert HF
  \rangle Z_{\sigma\rho\lambda\tau}\nonumber \\
&=&\Delta E_{int} 
   \sum_{\sigma\rho\lambda\tau} 
   \langle HF\vert
 [h_{\tau^\prime} e^\dagger_{\lambda^\prime} e_{\rho^\prime} e_{\sigma^\prime},
  e^\dagger_{\sigma} e^\dagger_{\rho} e_{\lambda} h^\dagger_{\tau}]
   \vert HF \rangle Z_{\sigma\rho\lambda\tau},
\end{eqnarray}

\end{widetext}

\noindent where $\Delta E_{int}$ is the excitation energy, measured with 
respect to the ground state of the $N_{e}$-electron system, and $H$ is 
the quantum dot Hamiltonian. These equations can be rewritten in a more 
compact form:

\begin{eqnarray}
\left( \matrix{A    & C \cr
               C^t  & B } \right)
\left( \matrix{V \cr Z}\right) =  \Delta E_{int}
\left( \matrix{V \cr Z}\right).
\label{matrix_p3h}
\end{eqnarray}

\noindent The matrix elements entering Eq. (\ref{matrix_p3h}) are computed 
by means of the following expressions:

\begin{widetext}

\begin{eqnarray}
\langle\sigma\tau\vert A\vert\sigma^\prime\tau^\prime\rangle\ =\ 
(\epsilon_{\sigma}+\epsilon_{\tau})\delta_{\sigma
\sigma^\prime}\delta_{\tau\tau^\prime}-\beta\langle\sigma^\prime\tau^\prime\vert \frac{1}{r_{eh}}
\vert\sigma\tau\rangle,
\label{sector_A}
\end{eqnarray}

\begin{eqnarray}
\langle\sigma\rho\lambda\tau\vert B\vert\sigma^\prime\rho^\prime\lambda^\prime\tau^\prime\rangle
\ &=&\ (\epsilon_{\sigma}+\epsilon_{\rho}+\epsilon_{\tau}-\epsilon_{\lambda})
\delta_{\sigma\sigma^\prime}\delta_{\rho\rho^\prime}\delta_{\tau\tau^\prime}
\delta_{\lambda\lambda^\prime} \nonumber\\
&-&\ \beta\left(\langle\rho^\prime\sigma^\prime\vert\frac{1}{r_{e}}\vert\overline{\rho\sigma}\rangle
\delta_{\tau\tau^\prime}\delta_{\lambda\lambda^\prime}+
\langle\lambda\rho^\prime\vert \frac{1}{r_{e}}\vert\overline{\rho\lambda^{\prime}}\rangle
\delta_{\tau\tau^\prime}\delta_{\sigma\sigma^\prime}+
\langle\lambda\sigma^\prime\vert \frac{1}{r_{e}}\vert\overline{\sigma\lambda^{\prime}}\rangle
\delta_{\tau\tau^\prime}\delta_{\rho\rho^\prime} \right. \nonumber \\
 &+&\left.\langle\lambda\rho^\prime\vert \frac{1}{r_{e}}\vert\overline{\lambda^{\prime}\sigma}\rangle
\delta_{\tau\tau^\prime}\delta_{\rho\sigma^\prime}+ 
\langle\lambda\sigma^\prime\vert\frac{1}{r_{e}}\vert\overline{\lambda^{\prime}\rho}\rangle
\delta_{\tau\tau^\prime}\delta_{\sigma\rho^\prime} \right) \ -\ 
\beta\left(\langle\sigma^\prime\tau^\prime\vert \frac{1}{r_{eh}}\vert\sigma\tau\rangle
\delta_{\rho\rho^\prime}\delta_{\lambda\lambda^\prime} \right. \nonumber \\
 &-&\left.\langle\rho^\prime\tau^\prime \vert\frac{1}{r_{eh}}\vert\sigma\tau\rangle
\delta_{\rho\sigma^\prime}\delta_{\lambda\lambda^\prime}-
\langle\lambda\tau^\prime\vert \frac{1}{r_{eh}}\vert\lambda^{\prime}\tau\rangle 
\delta_{\rho\rho^\prime}\delta_{\sigma\sigma^\prime}-
\langle\sigma^{\prime}\tau^\prime\vert \frac{1}{r_{eh}}\vert\rho\tau\rangle
\delta_{\sigma\rho^\prime}\delta_{\lambda\lambda^\prime} \right. \nonumber \\
 &+&\left.\langle\rho^{\prime}\tau^\prime\vert \frac{1}{r_{eh}}\vert\rho\tau\rangle
\delta_{\sigma\sigma^\prime}\delta_{\lambda\lambda^\prime} \right), 
\label{sector_B}
\end{eqnarray}

\begin{eqnarray}
\langle\sigma\rho\lambda\tau\vert C\vert\sigma^\prime\tau^\prime\rangle\ =\ 
\beta\left(
\langle\sigma^\prime\lambda\vert\frac{1}{r_{e}}
\vert\overline{\rho\sigma}\rangle\delta_{\tau\tau^\prime}-
\langle\lambda\tau^\prime\vert \frac{1}{r_{eh}}
\vert\sigma\tau\rangle\delta_{\rho\sigma^\prime}+
\langle\lambda\tau^\prime\vert \frac{1}{r_{eh}}
\vert\rho\tau\rangle\delta_{\sigma\sigma^\prime}\right),
\label{sector_C}
\end{eqnarray}

\begin{eqnarray}
\langle\sigma\tau\vert C^{t}\vert\sigma^\prime\rho^\prime\lambda^\prime\tau^\prime \rangle\ =\ \beta\left(
\langle\rho^\prime\sigma^\prime\vert\frac{1}{r_{e}}
\vert\overline{\sigma\lambda^\prime}\rangle\delta_{\tau\tau^\prime}-
\langle\sigma^\prime\tau^\prime\vert \frac{1}{r_{eh}}
\vert\lambda^\prime\tau\rangle\delta_{\sigma\rho^\prime}+
\langle\rho^\prime\tau^\prime\vert \frac{1}{r_{eh}}
\vert\lambda^\prime\tau\rangle\delta_{\sigma\sigma^\prime}\right),
\label{sector_CPrime}
\end{eqnarray}

\end{widetext}

\noindent where we used the notation $\langle ij\vert\vert\overline{kl}\rangle=
\langle ij\vert\vert kl\rangle-\langle ij\vert\vert lk\rangle$ for the Coulomb matrix elements, 
$\epsilon_{i}$ denotes the energy of the Hartree-Fock state $i$, and $\beta$ is the Coulomb 
interaction strength.

\end{document}